\documentclass[%
prx,
 amsmath,amssymb,
 aps,
twocolumn,
showkeys
]{revtex4-2}

\usepackage{graphicx}
\usepackage{epstopdf}
\usepackage{dcolumn}
\usepackage{bm}
\usepackage{color}
\usepackage[mathlines]{lineno}
\usepackage[breaklinks=true,colorlinks=true,linkcolor=blue,urlcolor=blue,citecolor=blue]{hyperref}
\usepackage{ragged2e}
\usepackage{enumitem}
\usepackage{bibunits}

\defaultbibliographystyle{elsarticle-num}
\defaultbibliography{biblio}

\newcommand{\lr}[1]{\left(#1\right)}
\newcommand{\lrs}[1]{\left[#1\right]}
\DeclareMathOperator{\tr}{tr}

\renewcommand\thefigure{\arabic{figure}}

\begin{document}


\title{Exit rights open complex pathways to cooperation}
\author{Chen Shen$^{1,2}$}
\thanks{Equal contribution}
\author{Marko Jusup$^2$}
\thanks{Equal contribution}
\author{Lei Shi$^1$}
\email{shi\_lei65@hotmail.com}
\author{Zhen Wang$^3$}
\email{w-zhen@nwpu.edu.cn}
\author{Matja{\v z} Perc$^{4,5,6}$}
\author{Petter Holme$^2$}

\affiliation{
\vspace{2mm}
\mbox{1. School of Statistics and Mathematics, Yunnan University of Finance and Economics, Kunming 650221, China}
\mbox{2. Tokyo Tech World Hub Research Initiative, Institute of Innovative Research,}
\mbox{Tokyo Institute of Technology, Tokyo 152-8550, Japan}
\mbox{3. Center for OPTical IMagery Analysis and Learning (OPTIMAL) and School of Mechanical Engineering,}
\mbox{Northwestern Polytechnical University, Xi'an 710072, China}
\mbox{4. Faculty of Natural Sciences and Mathematics, University of Maribor, 2000 Maribor, Slovenia}
\mbox{5. Department of Medical Research, China Medical University Hospital, China Medical University, Taichung 404, Taiwan}
\mbox{6. Complexity Science Hub Vienna, 1080 Vienna, Austria}
\vspace{2mm}
}

\date{\today}

\begin{abstract}
We study the evolutionary dynamics of the prisoner's dilemma game in which cooperators and defectors interact with another actor type called exiters. Rather than being exploited by defectors, exiters exit the game in favour of a small payoff. We find that this simple extension of the game allows cooperation to flourish in well-mixed populations when iterations or reputation are added. In networked populations, however, the exit option is less conducive to cooperation. Instead, it enables the coexistence of cooperators, defectors, and exiters through cyclic dominance. Other outcomes are also possible as the exit payoff increases or the network structure changes, including network-wide oscillations in actor abundances that may cause the extinction of exiters and the domination of defectors, although game parameters should favour exiting. The complex dynamics that emerges in the wake of a simple option to exit the game implies that nuances matter even if our analyses are restricted to incentives for rational behaviour.
\end{abstract}


\maketitle

\begin{bibunit}


In economic game theory, the conditions and consequences of quitting a game~\cite{solan2003quitting}, and voluntary participation in general, are fundamental topics~\cite{myerson}. In the theory of the evolution of cooperation, however, they are rarer guests~\cite{szabo2002evolutionary, ginsberg2019evolution}. Because evolutionary game theory traditionally concerns the competition between species, it is not surprising that the primary focus is on involuntary interactions~\cite{hofbauer1998evolutionary, mcnamara2020game}. Nevertheless, there is an increasing interest in modelling the interface between cooperation and social behaviour in human populations. We will also take this route and extend the canonical model of cooperation between selfish individuals -- the prisoner's dilemma~\cite{axelrod1981evolution, myerson, jackson2010social} -- with an option of exiting the game. To more realistically incorporate sociality, our players, or actors, will interact over model social networks~\cite{jackson2010social, szabo2007evolutionary, wang2015evolutionary}.

There are historical examples where the option to exit a game could have had a dramatic impact on the outcome. In the final years of the 1950s, China carried out far-reaching collectivisation of its society. Everyone in the countryside had to belong to a `people's commune' where people shared everything -- farming tools, seeding crops, draft animals, kitchens, and health care. Even private cooking was banned and replaced by communal canteens. Between 1958 and 1962, one of the worst famines in the history of humanity struck the country~\cite{ashton1992famine}. Ever since then, scholars have debated the connections between these social changes and the famine~\cite{meng2015institutional}.

One intriguing theory was proposed in 1990 by the economist Justin Yifu Lin of Peking University~\cite{lin1990collectivization}. He pointed out that with the establishment of the people's communes, leaving a collective was no longer an option. He reasoned that this revocation of the right to exit took away a disincentive to free ride, as now farmers could no longer avoid negative feedback loops of perfidy. Just how important this mechanism was in the onset of famine has been debated. For example, Refs.~\cite{dong1993does, macleod1993role} contend that Lin was wrong using various economic arguments, while Refs.~\cite{orbell1984cooperators, orbell1993social} lend support to the general idea of exit options promoting cooperation.

We will not dwell further on the question of how well Lin's hypothesis explains the connection between the collectivisation and famine. Instead, intrigued by this historical example, we will investigate in a more generic setting how much a simple right to exit can impact the evolution of cooperation. Our starting point is the prisoner's dilemma -- a basic mathematical formulation of the situation in which cooperation would be most beneficial in the long run, but only considering the next interaction, defection would be advantageous~\cite{myerson, jackson2010social, hofbauer1998evolutionary}. There are many mechanisms promoting cooperation in the prisoner's dilemma. Ref.~\cite{nowak2006five} divides these mechanisms into five categories -- kin and group selection, as well as direct, indirect, and network reciprocity. Others try to identify common principles behind all these mechanisms~\cite{eshel1982assortment, newton2018evolutionary, kay2020evolution}.

People interact in social networks~\cite{jackson2010social}. The structure of the networks can influence the game dynamics. Therefore, many authors have investigated games in which actors interact over model networks~\cite{szabo2007evolutionary, wang2015evolutionary}. We will investigate the prisoner's dilemma with an exit option on the regular lattice, as well as three additional types of network models: (i) small-world networks that have many triangles and short path-lengths characteristic of social networks~\cite{watts1998collective}, (ii) random regular graphs known to be very robust to perturbations, and (iii) scale-free networks that have fat-tailed degree distributions characteristic of socioeconomic systems~\cite{santos2005scale}. 

Networked populations have received tremendous attention among evolutionary game theorists upon the discovery that the Prisoner's Dilemma in lattices may generate spatial chaos~\cite{nowak1992evolutionary}. Exploring the role of network topology~\cite{abramson2001social}, and especially that of heterogeneous networks~\cite{holme2003prisoners}, has proven fruitful, leading up to a landmark result that when social networks are scale-free, cooperation dominates throughout much of the phase space of the Prisoner's Dilemma and other common social-dilemma games~\cite{santos2005scale, santos2006evolutionary}. Even in the multiplayer generalisation of the Prisoner's Dilemma called the Public Goods Game~\cite{hauert2003prisoner}, cooperation has been shown to benefit if the game is played in heterogeneous networks~\cite{santos2008social}. Although networked populations promote cooperation without adding strategic complexity, there have been numerous studies that extend the Prisoner's Dilemma in networks with, e.g., punishment~\cite{wang2013impact}, reward~\cite{wang2014rewarding}, reputation~\cite{wang2012inferring}, etc. Interestingly, empirical studies have failed to confirm some of these theoretical results. A scale-free topology, for example, was unable to promote cooperation among human actors above the levels established in a lattice~\cite{gracia2012heterogeneous}. Similarly, introducing peer punishment into simple networks of human actors left cooperation levels unchanged, all the while diminishing other benefits of network reciprocity~\cite{li2018punishment}. Empirical studies have -- unrelated to networks -- been known to produce conflicting or surprising results. Peer punishment thus may~\cite{dreber2008winners} or may not~\cite{wu2009costly, wu2009costly} promote cooperation, whereas rewarding may do so, but in a convoluted manner of exploiting a known cognitive bias~\cite{wang2018exploiting}.

Empirical studies notwithstanding, networked populations remain a pillar of modern evolutionary game theory. We build on this pillar by introducing a simple exit option that guarantees a small payoff to an exiter irrespective of what other actors do. Such a small payoff should intuitively be understood in the context of our motivational example on farming collectives, in which the farmer who exits their collective forgoes a larger potential benefit (i.e., the economies of scale), but still benefits from cultivating own land. We begin our analysis with a well-mixed population in which both one-shot and iterated prisoner's dilemma games with an exit option are played. We thereafter progressively add more complexity by considering populations in a lattice formation, as well as homogeneous and heterogeneous networks. In doing so, we observe a multitude of dynamic phenomena ranging from cyclic dominance to global oscillations to hub-node stabilisation.

\begin{table}
\caption{\label{t01} Payoff matrix for the weak prisoner's dilemma with an exit option.}
\begin{ruledtabular}
\begin{tabular}{cccc}
~   & $C$        & $D$        & $E$        \\
\hline
$C$ & 1          & 0          & 0          \\
$D$ & $b$        & 0          & 0          \\
$E$ & $\epsilon$ & $\epsilon$ & $\epsilon$ \\
\end{tabular}
\end{ruledtabular}
\begin{minipage}{0.48\textwidth}
\justify
The first row indicates that when a cooperator, $C$, meets another cooperator, defector $D$, or exiter $E$, they earn a payoff equal to one, zero, or zero, respectively. Analogously, when a defector meets a cooperator, defector, or exiter, they earn a payoff equal to $b\in \left(1,2\right]$, zero, or zero, respectively. Finally, exiters earn a payoff equal to $\epsilon\in \left[0,1\right)$, irrespective of whom they meet. In the most general variant of the prisoner's dilemma, a cooperator meeting another cooperator would earn the payoff $R$, a cooperator meeting a defector would earn $S$, a defector meeting a cooperator would earn $T$, and a defector meeting another defector would earn $P$, where these payoffs must satisfy $T>R>P>S$.
\end{minipage}
\end{table}

\section*{Methods}

The key elements of our modelling approach comprised (i) actions and payoffs, (ii) population structure, (iii) action selection, and (iv) simulation settings. We proceed to briefly describe each of these elements.

\paragraph*{Actions and payoffs.} For the sake of simplicity, we chose to base our model on the weak prisoner's dilemma~\cite{nowak1993spatial}. In this game, cooperators encountering cooperators receive the payoff equal to unity. Cooperators encountering defectors receive nothing. Conversely, defectors encountering cooperators receive the temptation payoff $b>1$. Defectors encountering defectors receive nothing. We added a third action to this setup, dubbed exit, such that exiters typically receive a small-but-positive payoff $\epsilon>0$ irrespective of whom they encounter. Cooperators and defectors encountering exiters receive nothing. Additional limits imposed on the payoffs were $b\leq2$ and $\epsilon<1$ in order to (i) make the cumulative value of mutual cooperation greater or equal to that of defection and (ii) make exiting less valuable than cooperating, respectively. The described setup is neatly summarised in Table~\ref{t01}.

\paragraph*{Population structure.} We assumed two general types of populations, well-mixed and networked. In the former case, an actor can encounter any other actor. In the latter case, an actor encounters only their neighbours as prescribed by the network. The basic network structure used in simulations was the regular lattice in two dimensions. Neighbourhood was von Neumann's, meaning that each actor has four neighbours: left, right, up, and down. Boundary conditions were periodic, meaning that in a lattice of size $L\times{}L$, actors in the $L$th row (column) are linked to actors in the first row (column). We also generated regular small-world networks and random regular networks by rewiring the underlying lattice, where the probability of rewiring any particular link ranged from 1\,\% (small-world) to 99\,\% (random). To generate scale-free networks for simulation purposes, we used the Barab\'{a}si-Albert algorithm~\cite{albert2002statistical}. All networks used in this study are visualised in Supplementary Information (SI) Fig.~S1.

\paragraph*{Action selection.} In well-mixed populations, action selection followed the usual replicator dynamics. Actors in networked populations selected their actions through imitation. Specifically, denoting the payoff earned by a focal actor $i$ with $\Pi_i$ and the payoff of a randomly selected neighbour $j$ with $\Pi_j$, the probability of the actor $i$ imitating the neighbour $j$ was given by the Fermi rule:
\begin{equation}
W_{i\leftarrow{}j}=\frac{1}{1+\exp\left(\frac{\Pi_i-\Pi_j}{K}\right)},
\label{e01}
\end{equation}
where $K$ measures the irrationality of selection. Note that as $K\rightarrow0$, the Fermi rule turns into the Heaviside step-function such that $W_{i\leftarrow{}j}=1$ if $\Pi_i<\Pi_j$, and $W_{i\leftarrow{}j}=0$ if $\Pi_i>\Pi_j$, while $W_{i\leftarrow{}j}=0.5$ if $\Pi_i=\Pi_j$ holds by definition. We set $K=0.1$ throughout the study.

\paragraph*{Simulation settings.} 

We arranged simulations in a series of Monte Carlo timesteps. In each timestep, we randomly selected a focal actor who then played the game with all their neighbours. We thereafter randomly selected one of the focal actor's neighbours, and allowed this neighbour to play the game with all their neighbours as well. We finally compared the payoffs of the focal actor and the selected neighbour to determine whether the focal actor imitates the neighbour or not.

We paid special attention in simulations to ensure that (i) transient dynamics had subsided and that (ii) finite-size effects had been eliminated. We thus ran simulation for \smash{$\mathcal{O}\left(10^4\right)$} timesteps, typically 50,000, while averaging actor abundances over the last \smash{$\mathcal{O}\left(10^3\right)$} timesteps, typically 5,000. Networks used in the study contained \smash{$\mathcal{O}\left(10^3\right)$} nodes, typically 5,000.

\section*{Results}

\paragraph*{Well-mixed populations.} We start our analysis from one of the simplest possible situations. Specifically, we consider a one-shot prisoner's dilemma with an exit option in a well-mixed population and, as mentioned, we simplify the exposition without much loss of generality by assuming the payoff structure of the weak prisoner's dilemma (Table~\ref{t01}). Under these conditions, only the monomorphic exiting equilibrium is stable, which is why the existence of the exit option is in no way helpful in establishing cooperation; see SI Remark 1. Actors simply choose to exit the game even if the payoff obtained by doing so is arbitrarily small; it is better to have some return with certainty than to risk getting exploited by defectors. 

The situation changes when we replace the single-shot game by an iterated game. Iterations, provided the game proceeds sufficiently many rounds, may favour cooperation; see SI Remark 2 and Ref.~\cite{axelrod1981evolution}. The exit option helps to eliminate defection irrespective of how small the exit payoff is. Put more technically, the equilibrium of full defection is unstable in this case and cannot be reached by means of evolutionary dynamics. Even if the population initially consisted of defectors alone, such a population would crumble under a slightest perturbation. The reason is that, provided the exit payoff is positive, exiting always confers more benefit than defecting. Actors ultimately choose to cooperate because, without defection as a viable option, cooperation is more beneficial than exiting the game. If we extend the game by adding a variable representing actor reputation, the effect is the same; see SI Remark 3. Our model thus shows that for well-mixed populations, the availability of the exit option supports cooperation, but only when accompanied by another mechanism, e.g., iterations (i.e., direct reciprocity) or reputation (i.e., indirect reciprocity), that makes cooperation feasible in the first place. These results open the question of what happens when the exit option is available in networked populations, in which the mechanism known as network reciprocity favours cooperation.

\begin{figure}[!t]
\includegraphics[scale=1.0]{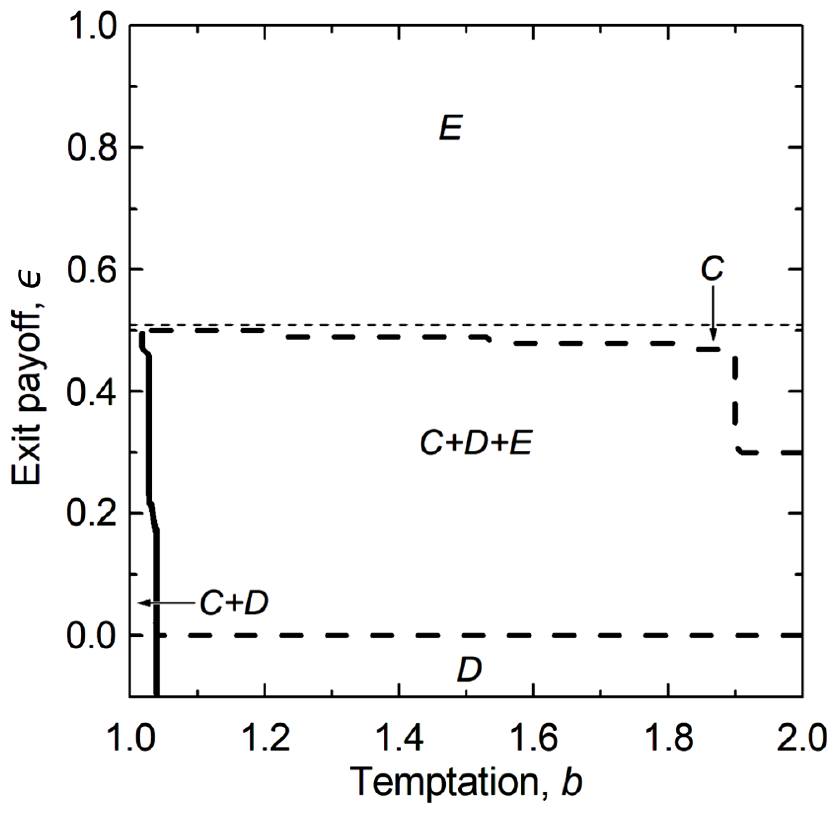}
\caption{\textbf{Cooperation is sustained, but rarely dominant, in networked populations with exit.} We plot the full $\epsilon$-$b$ phase diagram as obtained by Monte Carlo simulations of the weak prisoner's dilemma comprising an exit option in a lattice. When the exit option is highly rewarding, $\epsilon\gtrapprox0.52$, exiters dominate. A less rewarding exit option, $\epsilon\lessapprox0.52$, leads to four different outcomes. If temptation is small, $b\lessapprox1.04$, network reciprocity \textit{eo ipso} ensures that cooperators remain in the population indefinitely alongside defectors (the $C{+}D$ phase). Larger temptation values, $b\gtrapprox1.04$, lead to defector domination for $\epsilon\leq0$ (the $D$ phase), but otherwise sustain the coexistence of all three actor types (the $C+D+E$ phase) or lead to cooperator domination (the $C$ phase). Note that purely cooperative outcomes emerge only over a small domain of the $\epsilon$-$b$ phase plane.}
\label{f01}
\end{figure}

\begin{figure*}[!t]
\begin{minipage}{0.71\textwidth}
\raggedright\includegraphics[scale=1.0]{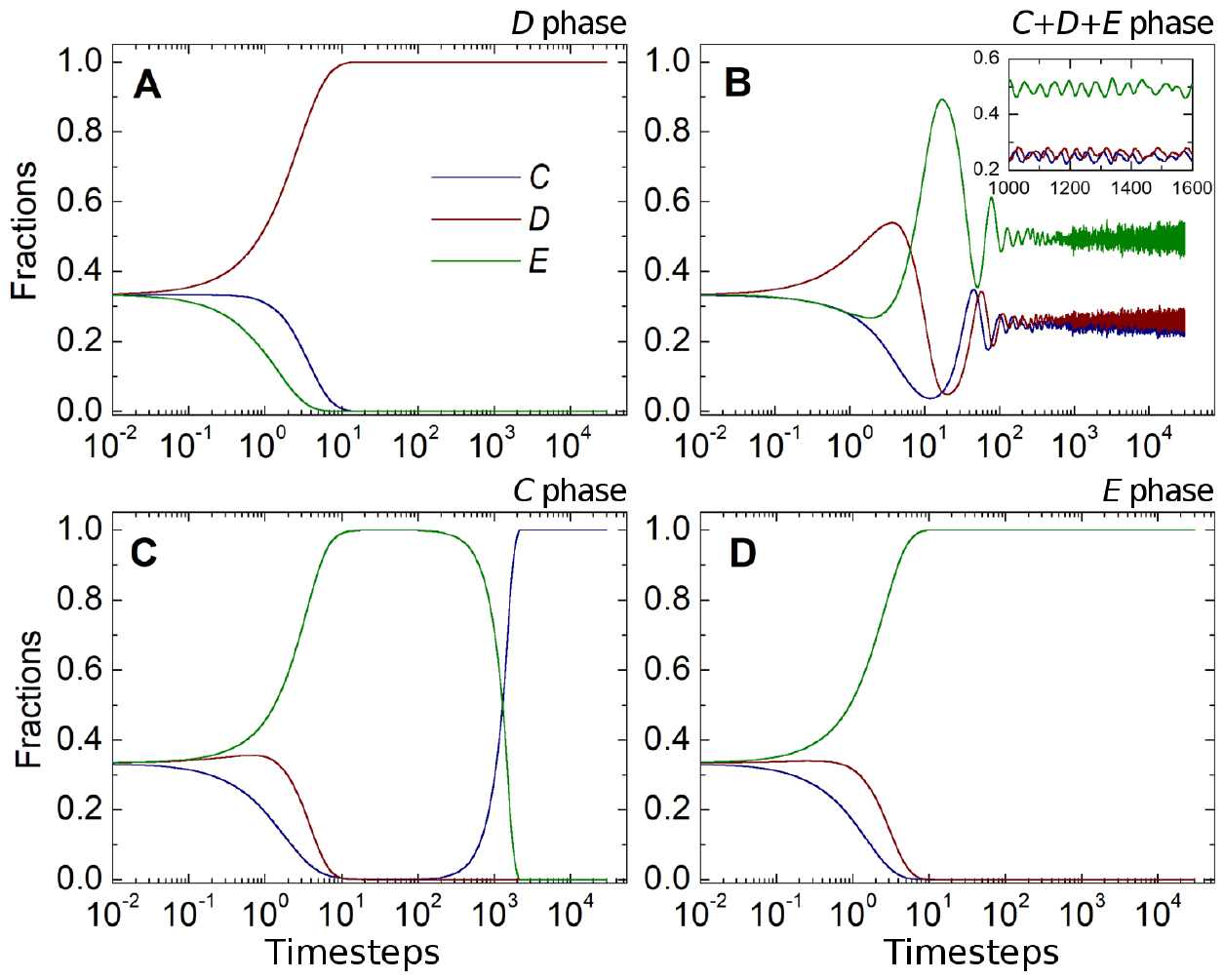}
\end{minipage}
\begin{minipage}[c][3.9in][t]{0.28\textwidth}
\caption{\textbf{Time dependence of actor abundances reveals cyclic dominance or sole game winners.} \textbf{A,} In the $D$ phase, defectors win by eliminating first exiters, then cooperators. \textbf{B,} In the $C{+}D{+}E$ phase, oscillating actor abundances are a signature of cyclic dominance. \textbf{C,} In the $C$ phase, the rise of exiters drives defectors to extinction, while rare cooperators survive and later prosper. There is, however, a narrow margin for this to happen. \textbf{D,} In the $E$ phase, the rise of exiters wipes out cooperators even before defectors. We show the results for over $10^4$ timesteps, which was sufficient for actor abundances to stabilise.}
\label{f02}
\end{minipage}
\end{figure*}

\paragraph*{Regular lattice.} To answer the question of how cooperation fares in networked populations with an exit option, we resorted to numerical Monte Carlo simulations; see Methods for details. We first performed simulations in lattices characterised by the von Neumann neighbourhood and the periodic boundary conditions. The game parameters were the payoffs $b$, $1<b\leq2$, and $\epsilon$, $-0.1<\epsilon<1$.

In Fig.~\ref{f01}, we show a phase diagram covering the full range of parameter values. We can see that adding an exit option can lead to complicated dynamics. First, when the exit payoff is $\epsilon\gtrapprox0.52$, exiters outcompete other actor types (the $E$ phase in Fig.~\ref{f01}). Conversely, when $\epsilon\lessapprox0.52$, there are four possible outcomes. Small temptation $b\lessapprox1.04$ allows network reciprocity alone to secure the coexistence of cooperators and defectors, while exiters get eliminated from the population (the $C{+}D$ phase in Fig.~\ref{f01}). Larger temptation $b\gtrapprox1.04$ gives rise either to (i) defector domination for $\epsilon\leq0$, (ii) the coexistence of all three actor types, or (iii) cooperator domination (respectively, the $D$ phase, the $C{+}D{+}E$ phase, and the $C$ phase in Fig.~\ref{f01}). A chief distinction between well-mixed and networked populations emerging from these results is that the latter permit dimorphic and trimorphic equilibria in which the different types of actors coexist. The exit option thus seems unable to entirely displace defection in networked populations, which is in contrast to our findings in well-mixed populations with either iterations or reputation, as described above. 

We can gain a better understanding of how the three types of actors affect one another by looking at the change of their abundances through time (Fig.~\ref{f02}). In the $D$ phase (Fig.~\ref{f02}A), exiters are the first to give way to defectors, followed shortly thereafter by cooperators. In the $C{+}D{+}E$ phase (Fig.~\ref{f02}B), it is cooperators who start giving way to defectors, but then -- with less cooperators around -- exiters temporarily outnumber defectors. Fewer defectors, in turn, allow cooperators to partly recover at the expense of exiters. This proceeds until recovering cooperators once more start giving way to defectors. We have thus described a phenomenon called cyclic dominance by which three actor types dominate one another in an intransitive manner. In our case, cooperators dominate exiters who dominate defectors who dominate cooperators. Cyclic dominance has proven influential in ecological~\cite{reichenbach2007mobility} and evolutionary game-theoretic~\cite{szolnoki2014cyclic} contexts, especially in voluntary dilemmas and extensions thereof~\cite{szabo2002evolutionary, szabo2002phase, guo2020novel}.

The phenomenon of cyclic dominance disappears in the $C$ phase (Fig.~\ref{f02}C) because here, a substantial rise of exiters drives defectors to extinction. At the same time, a tiny fraction of cooperators survive and, in the absence of defectors, eventually takes over the lattice. The rise of exiters in the $E$ phase (Fig.~\ref{f02}D), however, is so forceful that they wipe out cooperators even before defectors, thus remaining the sole actor type in the lattice.

\begin{figure*}[!t]
\begin{minipage}{0.71\textwidth}
\raggedright\includegraphics[scale=1.0]{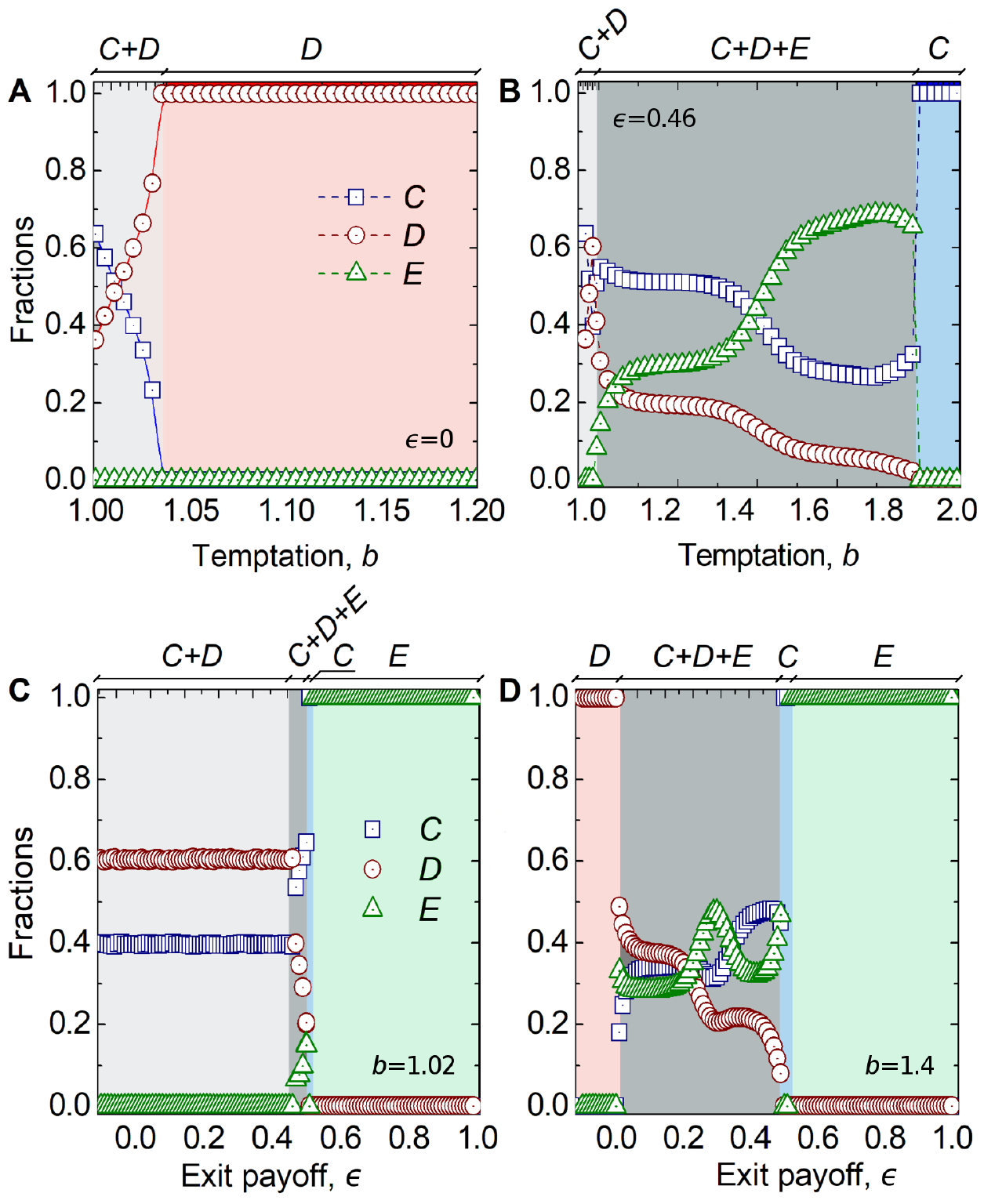}
\end{minipage}
\begin{minipage}[c][5.7in][t]{0.28\textwidth}
\caption{\textbf{Power relations between cooperators, defectors, and exiters exhibit intricate patterns.} \textbf{A,} Along the horizontal transect of the $\epsilon$-$b$ phase plane at $\epsilon=0$, network reciprocity alone is enough to secure the coexistence of cooperators and defectors for $b\lessapprox1.04$. Thereafter, defectors prevail. \textbf{B,} Along the horizontal transect at $\epsilon=0.46$, cooperators, defectors, and exiters coexist over the temptation range $1.04\lessapprox{}b\lessapprox1.90$ by way of cyclic dominance. For $b\gtrapprox1.90$, cooperators dominate. \textbf{C,} Along the vertical transect of the $\epsilon$-$b$ phase plane at $b=1.02$, network reciprocity secures the coexistence of cooperators and defectors up to a relatively large exit payoff of $\epsilon\approx0.45$. Between $0.45\lessapprox\epsilon\lessapprox0.50$, all three actor types coexist, whereas between $0.50\lessapprox\epsilon\lessapprox0.52$, there is a narrow strip of cooperator dominance. Thereafter, exiters prevail. \textbf{D,} Along the vertical transect at $b=1.4$, defectors dominate for $\epsilon\leq0$, the three actor types coexist for $0<\epsilon\lessapprox0.49$, cooperators dominate over a narrow strip between $0.49\lessapprox\epsilon\lessapprox0.52$, and finally, exiters dominate thereafter. Symbols (squares, circles, and triangles) indicate the average steady-state abundances of the three actor types.}
\label{f03}
\end{minipage}
\end{figure*}

\begin{figure*}[!t]
\includegraphics[scale=1.0]{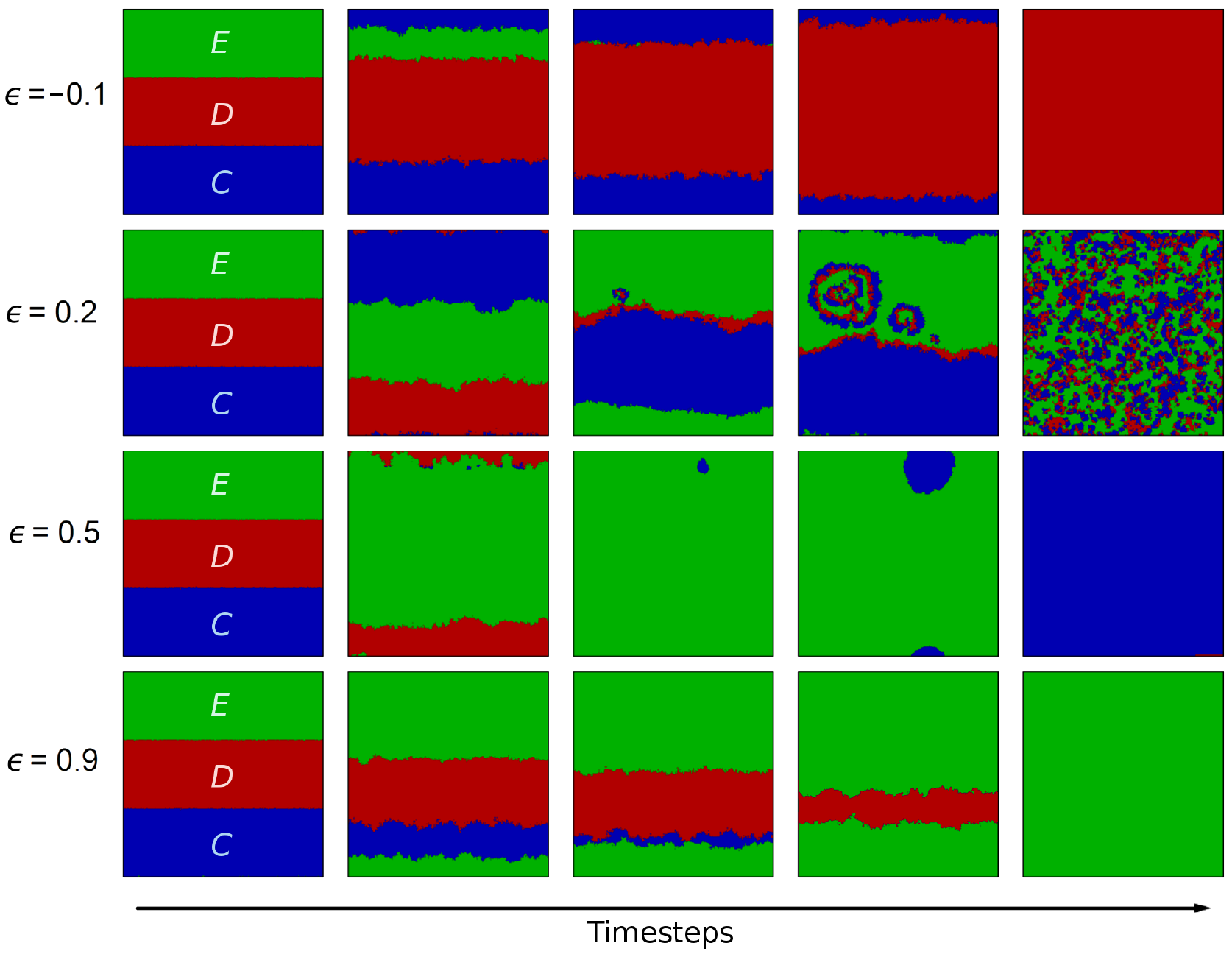}
\caption{\textbf{Snapshots of evolutionary dynamics expose in detail the interactions between actor types.} When the exit payoff is negative (top row), both cooperators and defectors oust exiters, who get eliminated first. Afterwards, defectors prevail. When the exit payoff is small-but-positive (second row), cyclic dominance ensues as recognisable by the eventual patchy distribution of actor types. A larger positive exit payoff (third row) enables exiters to eliminate defectors, but in the struggle between remaining cooperators and exiters, the former prevails. When the positive payoff is even larger (bottom row), defectors and exiters ousts cooperators. Defectors ultimately lose to exiters who dominate alone. All simulations were run with temptation $b=1.9$ for 30,000 timesteps to generate the final snapshots (rightmost column). The intermediate snapshots (second to fourth columns) were taken at different timesteps across rows to make the figure as illustrative as possible.}
\label{f04}
\end{figure*}

The relationships between actor types described here could be seen as power relations, in the sense of who dominates whom and in which conditions. In Fig.~\ref{f03}, we further analyse such relations by examining the equilibrium abundances of cooperators, defectors, and exiters along several transects of the phase space. These horizontal transects reveal power relations between the three actor types depending on the temptation payoff, $b$. In the usual weak prisoner's dilemma without exit, this payoff is equivalent to dilemma strength~\cite{wang2015universal} and thus a crucial determinant of the game outcome. Here, we find that when exiting is neutral or costly ($\epsilon\leq0$), network reciprocity can support cooperation by itself for small values of temptation ($b\lessapprox1.04$), but generally, defectors dominate (Fig.~\ref{f03}A). When, in contrast, exiting is marginally to moderately profitable ($0<\epsilon\lessapprox0.52$), network reciprocity still supports cooperation for small values of temptation ($b\lessapprox1.04$), but otherwise the domination of defectors is replaced by the coexistence of all three actor types (Fig.~\ref{f03}B).

The coexisting state is unusual in that the abundance of exiters increases with temptation, first more at the expense of defectors and later of cooperators. Temptation thus fails to entice defection but instead pushes actors to exit the game. This ultimately hurts defectors who can even go extinct by temptation being too large ($b\gtrapprox1.90$) and exiting sufficiently profitable ($0.30\lessapprox\epsilon\lessapprox0.52$). Without defectors to exploit them, cooperators become free to dominate (Fig.~\ref{f03}B).

In a similar vein as horizontal transects, vertical transects of the $\epsilon$-$b$ phase plane also reveal power relations between the three actor types, but this time depending on the exit reward $\epsilon$. For small temptation values, $b\lessapprox 1.04$, network reciprocity is enough to ensure the coexistence of cooperators and defectors for a relatively wide range of exit reward values (Fig.~\ref{f03}C). After crossing $\epsilon\approx0.45$, exiters are able to reduce the abundance of defectors, and after crossing $\epsilon\approx0.50$, defectors are eliminated, thus allowing cooperators to flourish (Fig.~\ref{f03}C). Cooperator domination, however, is short-lived because already beyond $\epsilon\approx0.52$, cooperators die out ahead of defectors, so exiters ultimately prevail (Fig.~\ref{f03}C). For larger temptation ($1.04\lessapprox{}b\lessapprox1.90$), the described situation partly repeats, that is, for $\epsilon\approx0.49$ there is again a narrow strip of cooperator domination, followed by a region of exiter domination (Fig.~\ref{f03}D). The situation changes below $\epsilon\approx0.49$ because network reciprocity is replaced by cyclic dominance, which ensures the coexistence of all three actor types between $0<\epsilon\lessapprox0.49$. Defectors prevail if $\epsilon\leq0$ (Fig.~\ref{f03}D).

Cyclic dominance gives rise to non-trivial dependence of actor abundances on the exit payoff. The average steady-state abundance of exiters thus first goes up at the expense of defectors from $\epsilon\approx0.21$, then goes down in favour of cooperators from $\epsilon\approx0.30$, only to go up one more time at the expense of defectors from $\epsilon\approx0.44$ (Fig.~\ref{f03}D). These undulations in exiter prevalence show that whenever exiting suppresses defection, then cooperation will soon increase. The margin for cooperator domination is generally narrow and widens only for the largest temptation that we consider $1.90\lessapprox b\leq2$ (see the $C$-phase in Fig.~\ref{f01}).

\begin{figure*}[!t]
\includegraphics[scale=1.0]{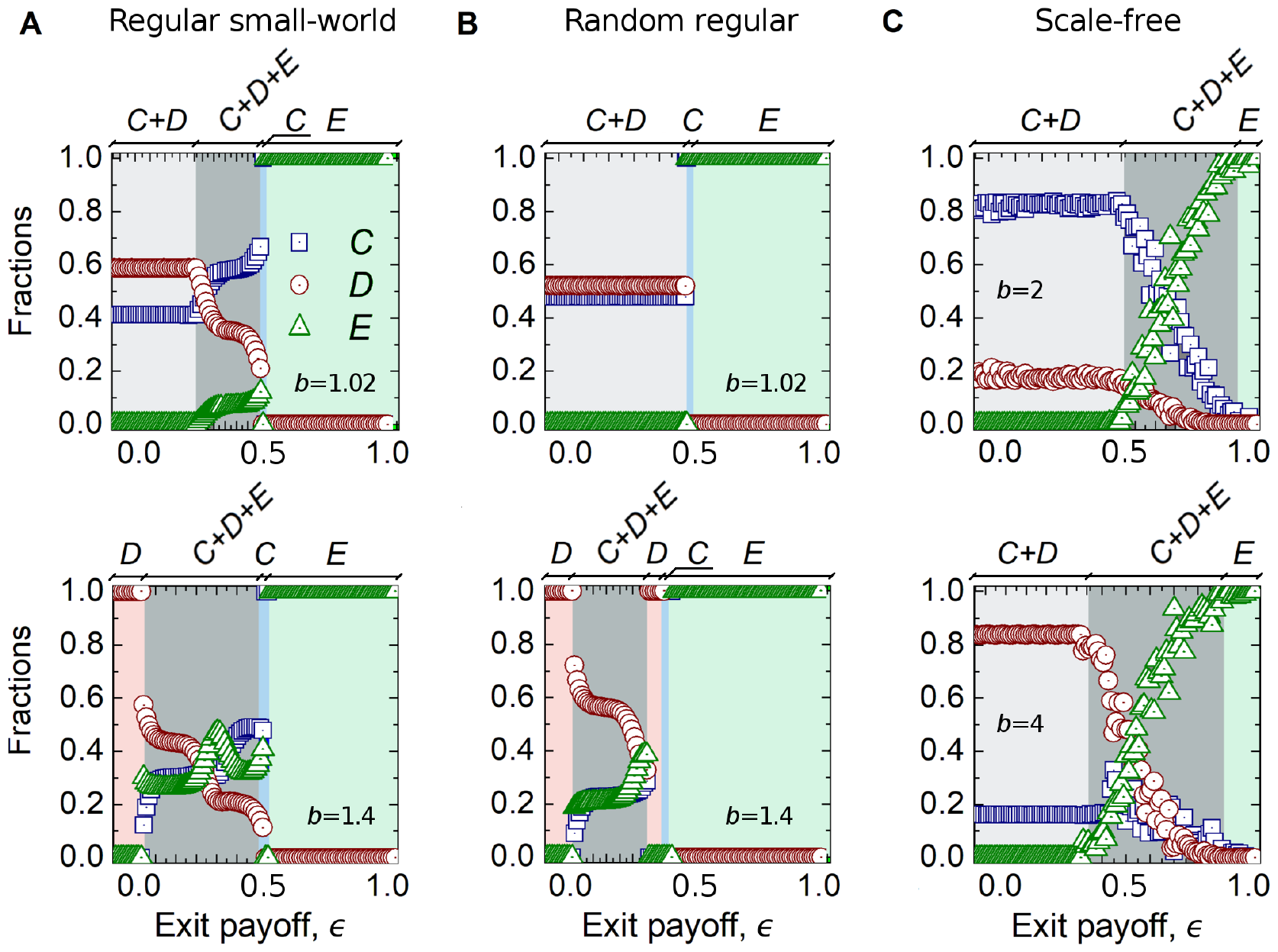}
\caption{\textbf{Network structure is an important determinant of evolutionary dynamics in the studied game.} \textbf{A,} Regular small-world networks have a smaller diameter, but otherwise remain similar to the regular lattice from which they were constructed. Consequently, the results here resemble those in Fig.~\ref{f03}C,\,D for both small temptation ($b=1.02$) in the upper panel and larger temptation ($b=1.4$) in the lower panel. \textbf{B,} Random regular networks differ more extensively from the regular lattice than regular small-world networks. This affects evolutionary dynamics. For example, there is no cyclic dominance in the upper panel when temptation is small ($b=1.02$). Cyclic dominance is seen again in the lower panel when temptation is larger ($b=1.4$), but here the rise of exiters can trigger defector domination for exit payoffs between $0.30\lessapprox\epsilon\lessapprox0.35$. \textbf{C,} Scale-free networks, constructed using the Barab\'{a}si-Albert algorithm~\cite{albert2002statistical}, give rise to fundamentally different evolutionary dynamics compared to other network structures. These networks support a large cooperator abundance even at temptation $b=2$, as seen in the upper panel. Some level of cooperation is possible even at temptation $b=4$, as seen in the lower panel. The coexistence of all three actor types, which arises at relatively large values of the exit payoff, is supported by a mechanism different from cyclic dominance. Finally, the sole domination of exiters is seen only at the near-maximum values of the exit payoff. Symbols (squares, circles, and triangles) indicate the average steady-state abundances of the three actor types.}
\label{f05}
\end{figure*}

In contrast to the time-series in Fig.~\ref{f02}, which show the aggregate development of actor abundances along the temporal dimension, snapshots of evolutionary dynamics provide insights into the development of local actor abundances along both spatial and temporal dimensions (Fig.~\ref{f04}). Snapshots thus open up the opportunity to reexamine the described phenomena from a microscopic perspective. Fixing temptation to $b=1.9$, we learn that non-positive exit payoffs make exiters weaker than cooperators or defectors (top row in Fig.~\ref{f04}). Consequently, cooperators and defectors jointly eliminate exiters, after which cooperators succumb to defectors (top row in Fig.~\ref{f04}). This sequence of events no longer transpires when the exit payoff turns positive. Then, instead, all three actor types get perpetually stuck in a loop of cyclic dominance (second row in Fig.~\ref{f04}). Making the exit payoff even more positive allows small pockets of cooperators to survive until the elimination of defectors by exiters. Afterwards, cooperators dominate exiters (third row in Fig.~\ref{f04}). Finally, if the exit payoff becomes too large, then even cooperators cannot stand up to exiters. Pressured from both defectors and exiters, cooperators get eliminated first, while defectors experience the same fate shortly thereafter, leaving exiters to dominate alone (bottom row in Fig.~\ref{f04}).

The above analysis of evolutionary snapshots demonstrates that network reciprocity combines with the exit option differently than iterations and reputation do. The latter allow that an arbitrarily small-but-positive exit payoff undermines defection (SI Fig.~S2). In contrast, in combination with network reciprocity, cooperation primarily happens via the coexistence of all three actor types due to cyclic dominance. How general are these observations? To answer this question, we proceed to examine whether and how the underlying network structure affects evolutionary dynamics.

\begin{figure*}[!t]
\begin{minipage}{0.71\textwidth}
\raggedright\includegraphics[scale=1.0]{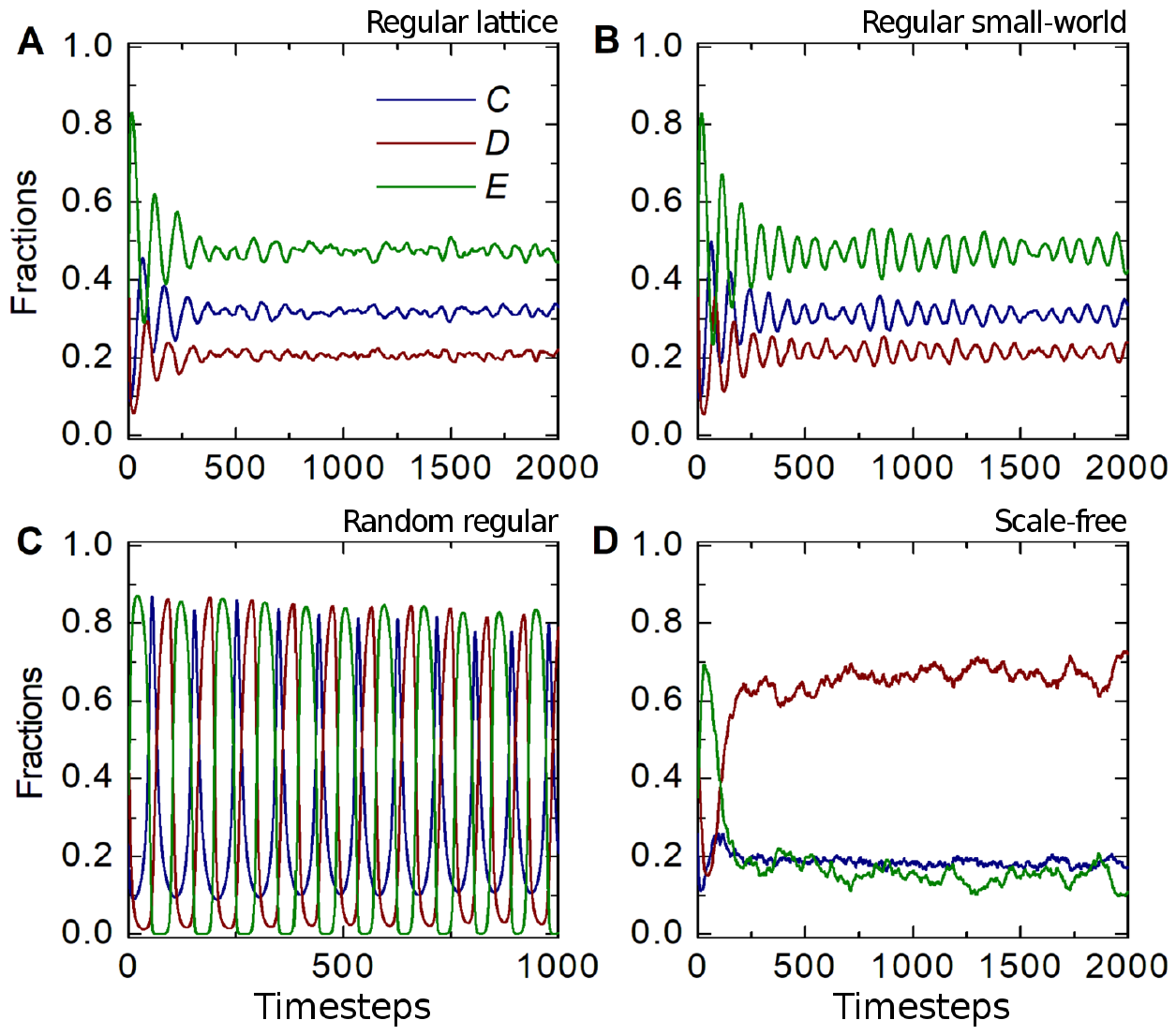}
\end{minipage}
\begin{minipage}[c][4.3in][t]{0.28\textwidth}
\caption{\textbf{Nature of cooperator, defector, and exiter coexistence changes with network structure.} \textbf{A,} In the regular lattice, initial large oscillations in average actor abundances quickly dampen and give way to much smaller oscillations that are a signature of cyclic dominance. \textbf{B,} Regular small-world networks have a smaller diameter than the regular lattice, but a similar density of squares, which somewhat increases the amplitude of oscillations. \textbf{C,} Random regular networks have not only a smaller diameter than the regular lattice, but also a much smaller density of squares. This is sufficient to trigger global-scale oscillations that may be large enough to eliminate exiters in some instances. \textbf{D,} In scale-free networks, hub nodes are predominantly cooperative, while small-degree nodes switch between defection and exiting. Temptation is $b=1.4$ and the exit payoff is $\epsilon=0.3$, except in the scale-free network where $b=4$ and $\epsilon=0.5$.}
\label{f06}
\end{minipage}
\end{figure*}

\paragraph*{Other networks.} To understand the effects of network structure on evolutionary dynamics, we ran simulations along the vertical transects of the $\epsilon$-$b$ phase plane in three additional network types: regular small-world, random regular, and scale-free (Fig.~\ref{f05}). The results of these simulations are thus analogous -- and best understood by comparing -- to the results in Fig.~\ref{f03}C,\,D. In constructing regular small-world networks, we started with the regular lattice and used random rewiring with the probability of 3\,\% to disconnect two neighbouring nodes and connect two nodes that had been distant before. This construction reduced the network diameter but left other properties, e.g., the density of squares, almost unchanged, which is why the simulation results for this network type and the regular lattice are similar (Fig.~\ref{f05}A). The only noteworthy difference is that for small temptation values ($b=1.02$), regular small-world networks support cyclic dominance more easily than the lattice (upper panel in Fig.~\ref{f05}A). For larger temptation values ($b=1.4$), we observe the same evolutionary dynamics in both network types (lower panel in Fig.~\ref{f05}A).

In constructing random regular networks, we followed the same procedure as for regular small-world networks, but with the rewiring probability as large as possible. This reduced not only the network diameter, but also the density of squares~\cite{holme2004structure}, causing evolutionary dynamics to change in two important ways (Fig.~\ref{f05}B). For small temptation ($b=1.02$), cyclic dominance vanishes (upper panel in Fig.~\ref{f05}B), whereas for larger temptation ($b=1.4$), there is a region of cyclic dominance as before, but now this region is separated from the narrow strip of cooperator domination by a strip of defector domination (lower panel in Fig.~\ref{f05}B). How it is possible that defectors become dominant  when the values of the exit payoff already strongly favour exiters? We will resolve this mystery shortly, after looking at the evolutionary dynamics in scale-free networks.

Scale-free networks, constructed using the Barab\'{a}si-Albert model~\cite{albert2002statistical}, lead to evolutionary dynamics that are fundamentally different than in other network types (Fig.~\ref{f05}C). Even if temptation is ramped up to $b=2$, network reciprocity supports a large cooperator abundance up to the exit payoff of $\epsilon\approx 0.48$ (upper panel in Fig.~\ref{f05}C). After that, the abundance of exiters increases linearly with the exit payoff up to $\epsilon\approx0.91$, when this actor type finally prevails. A similar picture holds even for temptation $b=4$, except that the abundances of cooperators and defectors switch places (lower panel in Fig.~\ref{f05}C).

It is illustrative at this point to look at the time-series of cooperator, defector, and exiter abundances when all three actor types coexist (Fig.~\ref{f06}). In the regular lattice, we find that initial large oscillations subside rather quickly, after which there are only small oscillations around the average abundances that are characteristic of cyclic dominance (Fig.~\ref{f06}A). The situation is similar in regular small-world networks, although the amplitude of oscillations around the average abundances is larger than before (Fig.~\ref{f06}B). The similarity between the time-series in these two cases seems to arise from almost the same density of squares in regular small-world networks as in the regular lattice. It would appear that squares keep oscillations local, and thereby small in amplitude (SI Fig.~S3A). This is perhaps expected for the regular lattice that lacks long-distance links, but less so for regular small-world networks that are much more compact.

Consistent with the above ideas, we further observe that as the density of squares approaches zero in random regular networks, oscillations become network-wide and develop very large amplitudes (Fig.~\ref{f06}C; see also SI Fig.~S3). There are even instances in which amplitudes are large enough to exterminate exiters, which is the reason why defector domination appears in the lower panel of Fig.~\ref{f05}B when the exit payoff should strongly support exiters. We also find that the nature of actor coexistence in scale-free networks is entirely different compared to other network structures. Hub nodes tend to cooperate, while small-degree nodes tend to switch between defection and exiting, which ultimately creates noisy rather than oscillating time-series of actor abundances (Fig.~\ref{f06}D). We visualise the described coexistence patterns by animated movies that are available at \url{doi.org/10.17605/OSF.IO/GRHSB}.

\section*{Discussion}

We have shown that adding an exceedingly simple exit option to a weak variant of the prisoner's dilemma is enough to generate complicated dynamics. In particular, we have seen that in well-mixed populations, an arbitrarily small-but-positive exit payoff, $\epsilon>0$, is sufficient to destabilise defection; see SI Remark 1. If there is also a viable cooperation-promoting mechanism in the form of iterations or reputation, cooperators can invade the population as long as their initial fraction is above $\epsilon$ (SI Fig.~S2B).

Combining the exit option with network reciprocity produces outcomes that differ greatly from those in well-mixed populations~\cite{hofbauer1998evolutionary}. We find that in networked populations an arbitrarily small-but-positive exit payoff typically leads to the coexistence of cooperators, defectors, and exiters through cyclic dominance. Coexistence by way of cyclic dominance is a subject of intense study in the contexts of biodiversity~\cite{frachebourg1996segregation, kerr2002local, reichenbach2007mobility, mobilia2010oscillatory, kelsic2015counteraction} and competition in microbial populations~\cite{durrett1997allelopathy, kirkup2004antibiotic, nahum2011evolution}. The results with more than three species in ecosystems, in fact, lend support to the conjecture that global oscillations are a general characteristic of realistic food webs \cite{rulquin2014globally}. Curiously, taking finite-size effects into account shows that cyclic dominance may in some instances compromise biodiversity and even cause extinction~\cite{reichenbach2006coexistence}. In the context of the evolution of cooperation, cyclic dominance often supports cooperation despite a large temptation to defect~\cite{szolnoki2010dynamically}, and in evolutionary games with more than three strategies, an important finding is that cyclic dominance provides an escape route from the negative impacts of antisocial punishment~\cite{szolnoki2017second}. For a detailed review of this extremely rich topic, see Ref.~\cite{szolnoki2014cyclic}. In the evolutionary game considered herein, it is particularly interesting that square-dense network structures, like the regular lattice, keep cyclic dominance local. In contrast, networks without squares, such as random regular networks, turn cyclic dominance into a global phenomenon. Dramatic oscillations may ensue (SI Fig.~S3), giving rise to sudden extinction of exiters and subsequent unexpected domination of defectors.

Parallels between our exiters and well-known loners, which rose to prominence as a mechanism behind cyclic dominance~\cite{szabo2002evolutionary, szabo2002phase, guo2020novel}, undoubtedly invite comparisons between the two. Loners are similar to exiters in that they opt out of the game to avoid getting exploited by defectors, but in doing so, loners differ from exiters in that they generate a small-but-positive payoff for the co-player in the game, regardless of whether this co-player is a cooperator, defector, or another loner. In this sense, the fact that exiters are just as responsible for cyclic dominance in networked populations as are loners shows that non-zero payoffs received by cooperators and defectors when interacting with loners are practically irrelevant for the observed dynamic phenomena. The exit option can thus be deemed, if not more basic, than at least more economical than the loner option. Exiters, furthermore, leave both cooperators and defectors completely hanging when they walk away from the game, which seems to correspond to various real-world situations. To exemplify, if completion of a scientific project rests on collaboration, two genuinely cooperative researchers should be able to complete the project as planned. If one of the researchers has free-riding tendencies, the project may still get completed, but the invested effort will be asymmetric. If, however, one researcher outright abandons the project for another project with a smaller-but-immediate payoff, the remaining researcher is left with little hope for success.

The view that exit option is beneficial for cooperation is being challenged by a new psychological study~\cite{haesevoets2019decision}. In their social-dilemma experiment, the authors find that with the introduction of exit rights, exiting replaces defection, which is in line with our model's predictions. This further leads to an increase in the relative cooperation frequency, where `relative' refers only to games without exiters. In absolute terms, however, the cooperation frequency decreases because many players choose to exit. In the model, whether more exiting leads to more or less cooperation depends on the exact setup. A larger temptation $b$ typically gives more power to defectors over cooperators, but with more defectors in the system, exiters also gain more space to spread, ultimately generating the situation in which more exiting is accompanied with less cooperation (e.g., Fig.~\ref{f03}B). A larger exit payoff $\epsilon$ mostly gives more power to exiters over defectors, but with more exiters in the system, cooperators also gain more space to spread, ultimately generating the situation in which more exiting is accompanied with more cooperation (e.g., Fig.~\ref{f05}A). It should not be forgotten here that, as the exit payoff becomes too large, exiters completely overpower the other two actor types. The authors of Ref.~\cite{haesevoets2019decision} conclude that ``both research and practice can gain greatly in richness by giving more consideration to exit options in the study of cooperation'', which is -- given the richness of our results -- a sentiment that we wholeheartedly agree with.

Exit rights have, to a degree, been studied at the interface between social and biological sciences. The most used methodology has been simulations with a focus on relative strategy effectiveness in iterated games~\cite{vanberg1992rationality, yamagishi1996selective, congleton2001help}. The results suggest that an exit option is beneficial for cooperation because exiting precludes exploitation by defectors. More recently, attention has turned towards social-dilemma experiments in which exiting has been realised through the ability to switch partners. The results also suggest that exiting benefits cooperation~\cite{rand2011dynamic, barclay2016partner}. Our conclusion is somewhat more nuanced -- while exit rights can help, they are certainly not a panacea.

Returning to the example of the Chinese famine, our results agree with Ref.~\cite{lin1990collectivization} in that having an exit option could save cooperation in the system. It is hard to interpret more of our results in that context; in conforming the model to networked evolutionary games, we lost the connection to that motivational example. Instead, we discovered that a seemingly minute adjustment to include exiters, leads to a plethora of dynamic phenomena. This shows that \emph{nuances matter} even if we restrict ourselves to the goal of economic and evolutionary game theory, that is, to elucidate incentives for rational behaviours. If we wanted to raise the bar and proceed to modelling general human behaviour~\cite{gintis2005behavioral, newton2018evolutionary}, details of the model would be even more important.

\section*{Article information}

\paragraph*{Acknowledgements.} This research was supported by the National Natural Science Foundation of China (grant no.~11931015) to L.\,S. We also acknowledge support from (i) the China Scholarship Council (scholarship no.~201908530225) to C.\,S., (ii) the Japan Society for the Promotion of Science (grant no.~20H04288) to M.\,J. as a co-investigator, (iii) the National Natural Science Foundation of China (grant no.~11671348) to L.\,S., (iv) the National Natural Science Foundation of China (grant no.~U1803263), the Thousand Talents Plan (grant no.~W099102), the Fundamental Research Funds for the Central Universities (grant no.~3102017jc03007), and the China Computer Federation--Tencent Open Fund (grant no.~IAGR20170119) to Z.\,W., (v) the Slovenian Research Agency (grant nos. J1-2457 and P1-0403) to M.\,P., and (vi) the Japan Society for the Promotion of Science (grant no.~18H01655) and the Sumitomo Foundation (grant for basic science research projects) to P.\,H.

\paragraph*{Author contributions.} C.\,S., M.\,J., L.\,S., and M.\,P. conceived research. C.\,S. and M.\,J. performed simulations. All co-authors discussed the results and wrote the manuscript.

\paragraph*{Conflict of interest.} Authors declare no conflict of interest.

\paragraph*{Code availability.} The code used in the study is freely available at \url{doi.org/10.17605/OSF.IO/GRHSB}.




\end{bibunit}

\clearpage
\onecolumngrid
\setcounter{equation}{0}
\renewcommand\theequation{S\arabic{equation}}
\setcounter{page}{1}
\renewcommand\thepage{A\arabic{page}}

\begin{bibunit}

\section*{
Supplementary Information for\\
``Exit rights open complex pathways to cooperation''}

\noindent\textbf{Remark 1.} We analyse, by means of the replicator equations, the evolutionary dynamics of a one-shot prisoner's dilemma with exit in a well-mixed population. Let $x$, $y$, and $z$ respectively denote the densities of cooperators ($C$), defectors ($D$), and exiters ($E$) in the population, where $0\leq{}x,y,z\leq{}1$ and $x+y+z=1$. The replicator equations are:
\begin{equation}
\begin{array}{l}
\dot{x} = x\lr{\Pi_C-\overline{\Pi}}, \\
\dot{y} = y\lr{\Pi_D-\overline{\Pi}}, \\
\dot{z} = z\lr{\Pi_E-\overline{\Pi}}.  
\end{array}
\label{eq11}
\end{equation}
The symbols $\Pi_C$, $\Pi_D$, and $\Pi_E$ denote, in that order, the expected payoff from cooperating, defecting, and exiting, whereas $\overline{\Pi}=x\Pi_C+y\Pi_D+z\Pi_E$ is the expected per-capita payoff of the whole population. Based on the payoffs defined in Table~\ref{t01} of the main text:
\begin{equation}
\begin{array}{l}
\Pi_C  = x, \\
\Pi_D  = bx, \\
\Pi_E  = \epsilon.
\end{array}
\label{eq12}
\end{equation}
Using the constraint $z=1-x-y$, we obtain:
\begin{equation}
\begin{array}{l}
\dot{x} = f\lr{x,y} = x\lrs{\lr{1-x}\lr{\Pi_C-\Pi_E}-y\lr{\Pi_D-\Pi_E}}=x\lrs{\lr{1-x}\lr{x-\epsilon}-y\lr{bx-\epsilon}}, \\
\dot{y} = g\lr{x,y} = y\lrs{\lr{1-y}\lr{\Pi_D-\Pi_E}-x\lr{\Pi_C-\Pi_E}}=y\lrs{\lr{1-y}\lr{bx-\epsilon}-x\lr{x-\epsilon}}, \\
\end{array}
\label{eq13}
\end{equation}
This system of equations has four equilibrium points: $\lr{1,0,0}$, $\lr{0,1,0}$, $\lr{0,0,1}$, and $\lr{\epsilon,0,1-\epsilon}$. To examine the stability of these equilibria, we calculate the Jacobian matrix:
\begin{equation}
J=\begin{bmatrix}
\frac{\partial{}f\lr{x,y}}{\partial{}x} & \frac{\partial{}f\lr{x,y}}{\partial{}y} \\
\frac{\partial{}g\lr{x,y}}{\partial{}x} & \frac{\partial{}g\lr{x,y}}{\partial{}y} \\
\end{bmatrix},
\label{eq14}
\end{equation}
where
\begin{equation}
\begin{array}{l}
\frac{\partial{}f\lr{x,y}}{\partial{}x} = x\lr{2-3x-2by}+\lr{-1+2x+y}\epsilon, \\
\frac{\partial{}f\lr{x,y}}{\partial{}y} = x\lr{-bx+\epsilon}, \\
\frac{\partial{}g\lr{x,y}}{\partial{}x} =
y\lr{b-2x-by+\epsilon}, \\
\frac{\partial{}g\lr{x,y}}{\partial{}y} = x\lr{b-x-2by}+\lr{-1+x+2y}\epsilon,
\end{array}
\label{eq15}
\end{equation}
and then look at the determinant and the trace of matrix $J$. The results of the stability analysis are:
\begin{itemize}[topsep=0pt,parsep=0pt,itemsep=0pt]
\item For point $\lr{1,0,0}$, we obtain $\det{}J=\lr{-1+b}\lr{-1+\epsilon}<0$, indicating that the equilibrium is unstable.
\item For point $\lr{0,1,0}$, we obtain $\det{}J=0$ and $\tr{}J=\epsilon\geq0$, again indicating that the equilibrium is unstable for any $0<\epsilon<1$.
\item For point $\lr{0,0,1}$, we obtain $\det{}J=\epsilon^2\geq0$ and $\tr{}J=-2\epsilon\leq0$, indicating that the equilibrium is stable for any $0<\epsilon<1$.
\item Finally, for point $\lr{\epsilon,0,1-\epsilon}$, we obtain $\det{}J=\epsilon^2\lr{-1+b}\lr{1-\epsilon}\geq0$ and $\tr{}J=\lr{b-\epsilon}\epsilon\geq0$, indicating one more time that the equilibrium is unstable.
\end{itemize}
In summary, all actors exit the game provided that for doing so they receive an arbitrarily small payoff.

\vspace{2mm}

\noindent\textbf{Remark 2.} We extend our previous analysis of the one-shot prisoner's dilemma with exit in a well-mixed population to a situation when the game is iterated. Iterations mean that two actors play the game for an unspecified number of rounds determined by a termination probability $q$. Precisely, the game may be terminated after each round with probability $q$, or it may continue with probability $1-q$. We furthermore assume that cooperative actors in the iterated game cooperate conditionally, that is, resort to the tit-for-tat strategy, because it is a well-known result of evolutionary game theory that unconditional cooperation is easily undermined by defection~\cite{nowak1998dynamics}. Cooperative actors thus start the game with cooperation, and proceed in the same fashion unless they are paired with a defector, in which case defection in the previous round is met with defection by the cooperative actor in the current round. Under such circumstances, the payoff matrix transforms into:
\begin{equation}
\begin{bmatrix}
\frac{1}{q}        & 0                  & 0                  \\
b                  & 0                  & 0                  \\
\frac{\epsilon}{q} & \frac{\epsilon}{q} & \frac{\epsilon}{q}
\end{bmatrix}.
\label{eq21}
\end{equation}
Based on this matrix, the expected payoffs from cooperating, defecting, and exiting become:
\begin{equation}
\begin{array}{l}
\Pi_C  = \frac{x}{q}, \\
\Pi_D  = bx, \\
\Pi_E  = \frac{\epsilon}{q}.
\end{array}
\label{eq22}
\end{equation}
Consequently, for the functions $f\lr{x,y}$ and $g\lr{x,y}$ as defined in Eq.~(\ref{eq13}), we get:
\begin{equation}
\begin{array}{l}
f\lr{x,y} = x\lrs{\lr{1-x}\frac{x-\epsilon}{q}-y\lr{bx-\frac{\epsilon}{q}}}, \\
g\lr{x,y} = y\lrs{\lr{1-y}\lr{bx-\frac{\epsilon}{q}}-x\frac{x-\epsilon}{q}}. \\
\end{array}
\label{eq23}
\end{equation}
The iterated game has the same four equilibrium points as the single-shot game: $\lr{1,0,0}$, $\lr{0,1,0}$, $\lr{0,0,1}$, and $\lr{\epsilon,0,1-\epsilon}$. To examine the stability of these equilibria, we calculate the elements of the Jacobian matrix (Eq.~\ref{eq14}):
\begin{equation}
\begin{array}{l}
\frac{\partial{}f\lr{x,y}}{\partial{}x} = \frac{x}{q}\lr{2-3x-2bqy}+\lr{-1+2x+y}\frac{\epsilon}{q}, \\
\frac{\partial{}f\lr{x,y}}{\partial{}y} = x\lr{-bx+\frac{\epsilon}{q}}, \\
\frac{\partial{}g\lr{x,y}}{\partial{}x} =
\frac{y}{q}\lr{bq-2x-bqy+\epsilon}, \\
\frac{\partial{}g\lr{x,y}}{\partial{}y} = \frac{x}{q}\lr{bq-x-2bqy}+\lr{-1+x+2y}\frac{\epsilon}{q},
\end{array}
\label{eq24}
\end{equation}
and then look at the determinant and the trace of matrix $J$. The results of the stability analysis are:
\begin{itemize}[topsep=0pt,parsep=0pt,itemsep=0pt]
\item For point $\lr{1,0,0}$, we obtain \smash{$\det{}J=\frac{1}{q^2}\lr{-1+bq}\lr{-1+\epsilon}$} and \smash{$\tr{}J=\frac{1}{q}\lr{-2+bq+\epsilon}$}. The determinant is positive, while the trace is negative if $q<\frac{1}{b}$. A sufficiently low termination probability, therefore, makes the fully cooperative equilibrium stable.
\item For point $\lr{0,1,0}$, we obtain $\det{}J=0$ and \smash{$\tr{}J=\frac{\epsilon}{q}\geq0$}. Accordingly, irrespective of the termination probability, the fully defecting equilibrium is unstable when there is an arbitrarily small exit payoff.
\item For point $\lr{0,0,1}$, we obtain \smash{$\det{}J=\frac{\epsilon^2}{q^2}\geq0$} and \smash{$\tr{}J=-\frac{2\epsilon}{q}\leq0$}. This indicates that the fully exiting equilibrium is stable provided there is an arbitrarily small payoff associated with the exit option.
\item For point $\lr{\epsilon,0,1-\epsilon}$, we obtain \smash{$\det{}J=\frac{\epsilon^2}{q^2}\lr{-1+bq}\lr{1-\epsilon}$} and \smash{$\tr{}J=\lr{b-\frac{\epsilon}{q}}\epsilon$}. The determinant here can be positive if \smash{$q>\frac{1}{b}$}, but in that case the trace is also positive, indicating that the mixed cooperative-exiting equilibrium is unstable.
\end{itemize}
In summary, iterations may overcome the dilemma in favour of cooperation if the termination probability is sufficiently low. The exit option meanwhile eliminates defection regardless of how small the exit payoff is (Fig.~\ref{fS02}). In the mix of cooperators and exiters, the smaller the values of $q$ and $\epsilon$, the more likely it is for cooperation to prevail.

\vspace{2mm}

\noindent\textbf{Remark 3.} The success of iterations in promoting cooperation is due to a mechanism called direct reciprocity that manifests in the tit-for-tat strategy of conditional cooperators~\cite{taylor2007transforming}. The gist of direct reciprocity is that cooperative actors have the time to assess whether they are paired with cooperators and then act accordingly. This naturally leads to a question if there could be other, more indirect, ways to assess whether someone is a cooperator. It turns out that reputation precisely serves this purpose, leading to the evolution of cooperation through indirect reciprocity~\cite{nowak2005evolution}. Assuming that an actor's reputation is known with a probability $p$, the payoff matrix of a single-shot game transforms into:
\begin{equation}
\begin{bmatrix}
1         & 0        & 0         \\
\lr{1-p}b & 0        & 0         \\
\epsilon  & \epsilon & \epsilon
\end{bmatrix}.
\label{eq31}
\end{equation}
The expected payoffs from cooperating, defecting, and exiting then become:
\begin{equation}
\begin{array}{l}
\Pi_C  = x, \\
\Pi_D  = \lr{1-p}bx, \\
\Pi_E  = \epsilon,
\end{array}
\label{eq32}
\end{equation}
while the functions $f\lr{x,y}$ and $g\lr{x,y}$ as defined in Eq.~(\ref{eq13}) turn into:
\begin{equation}
\begin{array}{l}
f\lr{x,y} = x\lrs{\lr{1-x}\lr{x-\epsilon}-y\lr{\lr{1-p}bx-\epsilon}}, \\
g\lr{x,y} = y\lrs{\lr{1-y}\lr{\lr{1-p}bx-\epsilon}-x\lr{x-\epsilon}}. \\
\end{array}
\label{eq33}
\end{equation}
The single-shot game with indirect reciprocity has the same four equilibrium points as before: $\lr{1,0,0}$, $\lr{0,1,0}$, $\lr{0,0,1}$, and $\lr{\epsilon,0,1-\epsilon}$. To examine the stability of these equilibria, we calculate the elements of the Jacobian matrix (Eq.~\ref{eq14}):
\begin{equation}
\begin{array}{l}
\frac{\partial{}f\lr{x,y}}{\partial{}x} = -3x^2-\lr{1-y}\epsilon+2x\lrs{1-b\lr{1-p}y+\epsilon}, \\
\frac{\partial{}f\lr{x,y}}{\partial{}y} = x\lrs{-b\lr{1-p}x+\epsilon}, \\
\frac{\partial{}g\lr{x,y}}{\partial{}x} =
y\lrs{-2x + b\lr{1-p}\lr{1-y}+\epsilon}, \\
\frac{\partial{}g\lr{x,y}}{\partial{}y} = -x^2+b\lr{1-p}x\lr{1-2y}+\lr{-1+x+2y}\epsilon,
\end{array}
\label{eq34}
\end{equation}
and then look at the determinant and the trace of matrix $J$. The results of the stability analysis are:
\begin{itemize}[topsep=0pt,parsep=0pt,itemsep=0pt]
\item For point $\lr{1,0,0}$, we obtain $\det{}J=\lrs{1-b\lr{1-p}}\lr{1-\epsilon}$ and $\tr{}J=-2 + b\lr{1-p}+\epsilon$. The determinant is positive, while the trace is negative if $p>\frac{b-1}{b}$. A sufficiently high probability of knowing the other actor's reputation thus makes the fully cooperative equilibrium stable.
\item For point $\lr{0,1,0}$, we obtain $\det{}J=0$ and $\tr{}J=\epsilon\geq0$. Accordingly, irrespective of the probability of knowing the other actor's reputation, the fully defecting equilibrium is unstable when there is an arbitrarily small exit payoff.
\item For point $\lr{0,0,1}$, we obtain $\det{}J=\epsilon^2\geq0$ and $\tr{}J=-2\epsilon\leq0$. This indicates that the fully exiting equilibrium is stable provided there is an arbitrarily small payoff associated with the exit option.
\item For point $\lr{\epsilon,0,1-\epsilon}$, we obtain $\det{}J=\epsilon^2\lrs{-1+b\lr{1-p}}\lr{1-\epsilon}$ and $\tr{}J=\lrs{b\lr{1-p}-\epsilon}\epsilon$. The determinant here can be positive if \smash{$p>\frac{b-1}{b}$}, but in that case the trace is also positive, indicating that the mixed cooperative-exiting equilibrium is unstable.
\end{itemize}
In summary, indirect reciprocity may overcome the dilemma in favour of cooperation if the probability of knowing the other actor's reputation is sufficiently high. The exit option meanwhile eliminates defection regardless of how small the exit payoff is (Fig.~\ref{fS02}). In the mix of cooperators and exiters, the larger the value of $p$ and the smaller the value $\epsilon$, the more likely it is for cooperation to prevail.

\clearpage

\renewcommand\thefigure{S\arabic{figure}}
\setcounter{figure}{0}  
\section*{Supplementary Figures}

\vfill

\begin{figure}[!h]
\includegraphics[scale=1.0]{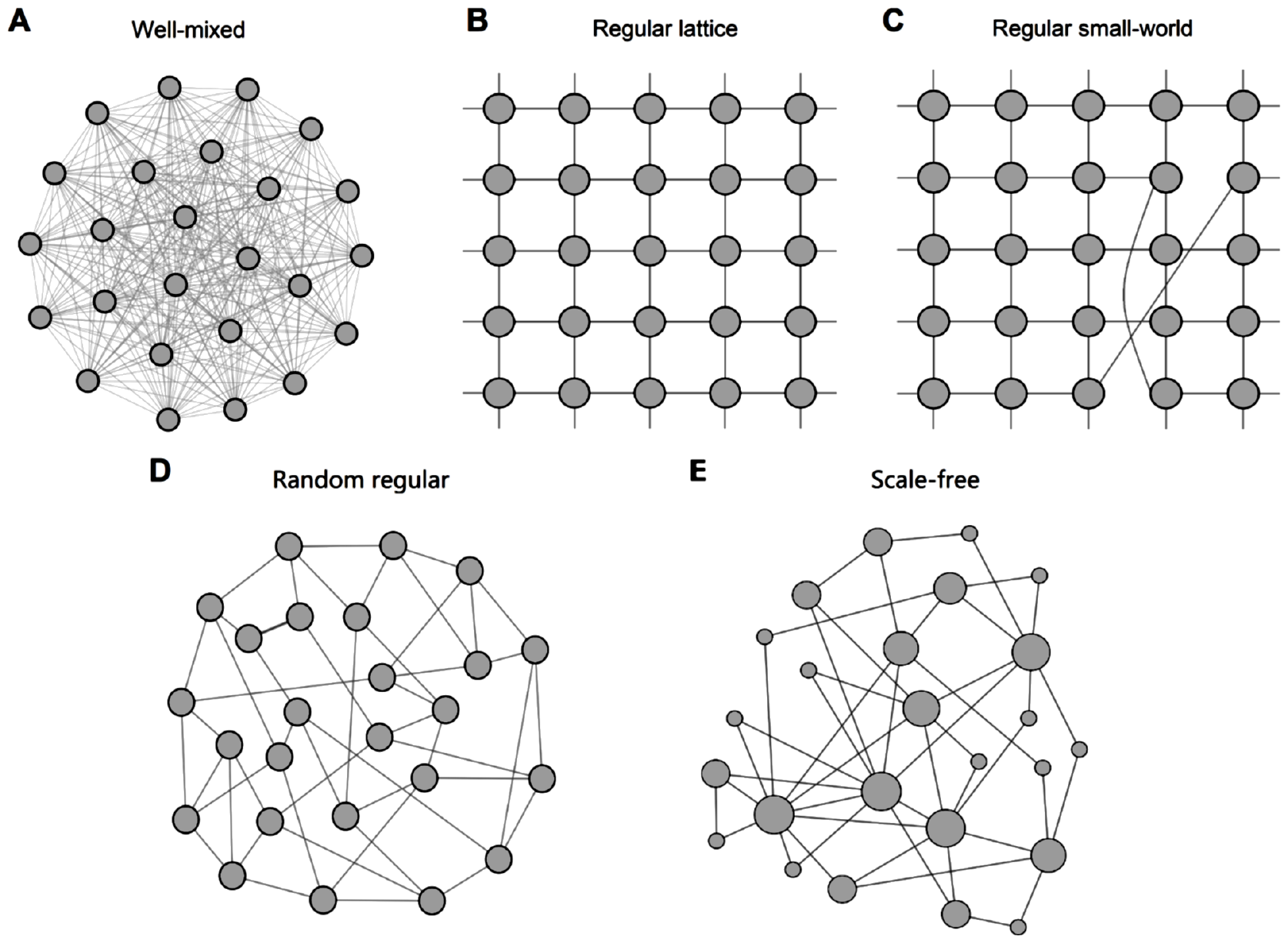}
\caption{\textbf{Network structures used in the study.} \textbf{A,} Well-mixed populations can be represented as fully connected networks because any two actors may play the game with one another at a given timestep. \textbf{B,} Regular lattices with the von Neumann neighbourhood and a periodic boundary are two-dimensional network structures in which every node has four neighbours (left, right, up, and down), and nodes at one end of the lattice link to the corresponding nodes at the other end (represented with the open links protruding from nodes in the first and fifth row and column). \textbf{C,} Regular small-world networks are created by rewiring a regular lattice such that two, randomly chosen, neighbouring nodes are disconnected and then connected to two nodes that were distant before. The rewiring probability is typically small, 3\,\% in our study. \textbf{D,} Random-regular networks are created following the same algorithm as for regular small-world networks, but with a large rewiring probability, 99\,\% in our study. \textbf{E,} Scale-free networks are characterised by a wide range of node degrees, where nodes with an exceptionally high degree are often called hub nodes.}
\label{fS01}
\end{figure}

\vfill

\clearpage
\phantom{Invisible text.}

\vfill

\begin{figure}[!h]
\includegraphics[scale=1.0]{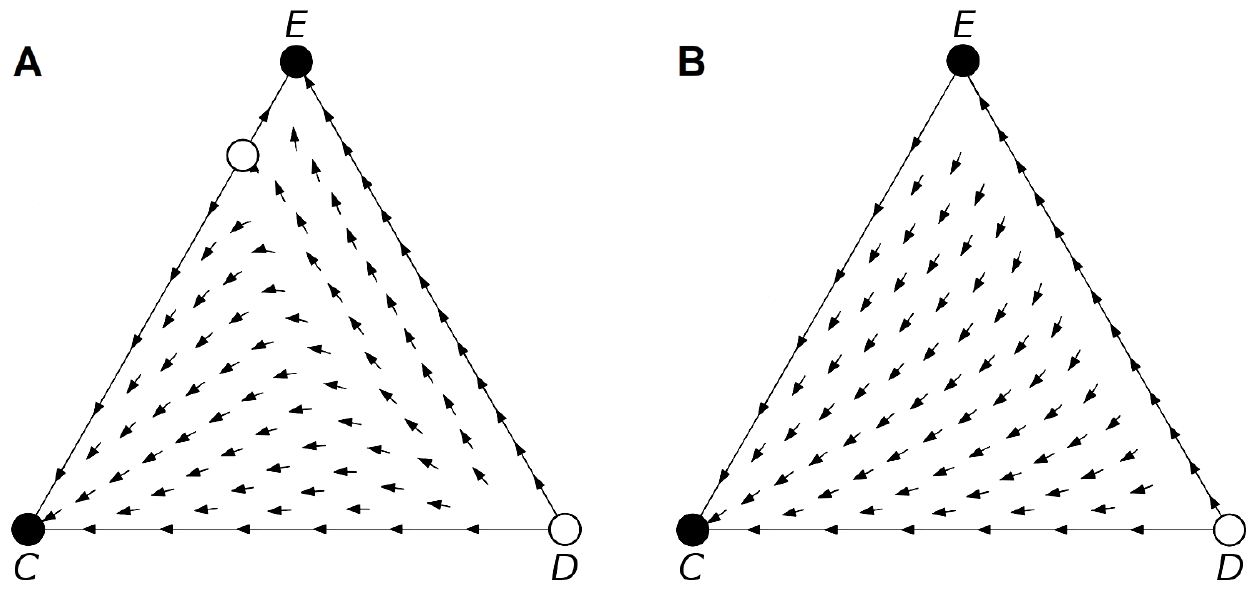}
\caption{\textbf{Exiting destabilises defection, while reciprocity ensures cooperation.} In well-mixed populations with direct or indirect reciprocity, there are three monomorphic evolutionary equilibria (cooperation, defection, and exiting respectively denoted $C$, $D$, and $E$) and one dimorphic equilibrium (cooperation-exiting). Any positive exit payoff, $\epsilon>0$, destabilises the $D$ equilibrium. In contrast, whether the $C$ equilibrium is stable or not depends on the strength of reciprocity, that is, the game termination probability under direct reciprocity or the probability of knowing the other actor's reputation under indirect reciprocity. \textbf{A,} The panel shows a ternary plot of evolutionary dynamics under indirect reciprocity when the exit payoff is $\epsilon=0.2$. The $C$ and $E$ monomorphic equilibria are stable (black disks), whereas the two other equilibria are unstable (white disks). Because of the relatively large exit payoff, evolutionary dynamics converge to the $E$ equilibrium when the initial abundance of cooperators is small. Otherwise, cooperation evolves. \textbf{B,} In the limit of an infinitesimally small exit payoff (here, $\epsilon=0.01$), the monomorphic $E$ equilibrium and the dimorphic equilibrium coincide. Cooperation evolves irrespective of how small the initial cooperator abundance is. In both ternary plots, the probability of knowing the other actor's reputation is $p=0.75$, whereas temptation is $b=1.2$. Analogous ternary plots can be obtained under direct reciprocity too.}
\label{fS02}
\end{figure}

\vfill

\clearpage
\phantom{Invisible text.}

\vfill

\begin{figure}[!h]
\includegraphics[scale=1.0]{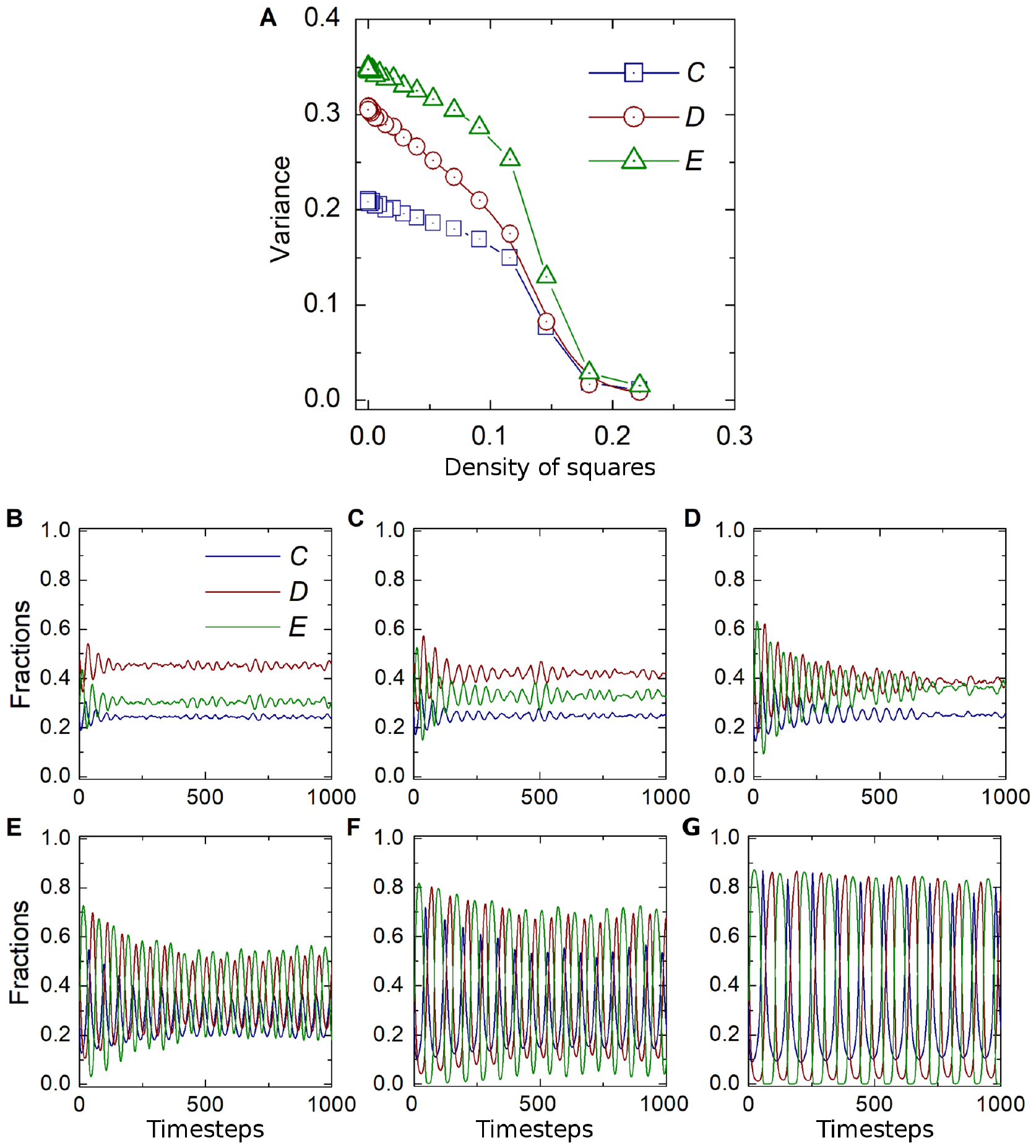}
\caption{\textbf{How local cyclic dominance turns into global oscillations of actor abundances.} \textbf{A,} The panel displays the variance of actor abundances (estimated over the last 5,000 simulation timesteps) as the functions of the density of squares, where this latter quantity was calculated following the definition in Ref.~\cite{holme2004structure}. Larger variances reflect larger oscillation amplitudes around equilibrium actor abundances. The density of squares decreases from the regular lattice to random regular networks by increasing the link rewiring probability from zero to unity. Here, temptation is $b=1.4$ and the exit payoff is $\epsilon=0.3$. \textbf{B--G,} In random regular networks (constructed from the regular lattice with the rewiring probability of 0.99), the oscillation amplitude of actor abundances is large for certain sets of parameter values. Here, temptation is $b=1.4$, while the exit payoff progressively increases from $\epsilon=0.25$ to $\epsilon=0.30$ in steps of 0.01. The exiter abundance oscillates with the largest amplitude which is why this actor type goes extinct when the exit payoff is between $0.30\lessapprox\epsilon\lessapprox0.35$.}
\label{fS03}
\end{figure}

\vfill

\clearpage


\end{bibunit}

\end{document}